\documentclass[reprint,showpacs,preprintnumbers,amsmath,amssymb,aip,nolongbibilography]{revtex4-1}
\usepackage{graphicx,wasysym}% Include figure files
\usepackage[usenames,dvipsnames]{color}
\usepackage{dcolumn}% Align table columns on decimal point
\usepackage{bm}% bold math
\usepackage{tabulary}
\usepackage{fancyhdr}

\newcommand{\Rp}{R_\mathrm{p}}
\newcommand{\Tm}{T_{\mathrm{m}}}
\newcommand{\Tg}{T_{\mathrm{g}}}
\newcommand{\Ts}{T_{\mathrm{s}}}
\newcommand{\To}{T_{\mathrm{o}}}
\newcommand{\Rt}{R}
\definecolor{light-gray}{gray}{0.45}
\begin{document}

\pagestyle{fancy}
\fancyhead{}
\renewcommand{\headrulewidth}{0.0pt}

%\fancyhead[CO,CE]{\textcolor{light-gray}{\bf{DRAFT}}}
\preprint{1}

\title{Phase diagram of supercooled water confined to hydrophilic nanopores}% Force line breaks with \\

\author{David T Limmer}
\author{David Chandler}%
 \email{chandler@cchem.berkeley.edu}
\affiliation{%
Department of Chemistry, University of California, Berkeley, USA 94609
}%

\date{\today}
\begin{abstract}
We present a phase diagram for water confined to cylindrical silica nanopores in terms of pressure, temperature and pore radius.  The confining cylindrical wall is hydrophilic and disordered, which has a destabilizing effect on ordered water structure.  The phase diagram for this class of systems is derived from general arguments, with parameters taken from experimental observations and computer simulations and with assumptions tested by computer simulation.  Phase space divides into three regions: a single liquid, a crystal-like solid, and glass.  For large pores, radii exceeding 1 nm, water exhibits liquid and crystal-like behaviors, with abrupt crossovers between these regimes.   For small pore radii, crystal-like behavior is unstable and water remains amorphous for all non-zero temperatures.  At low enough temperatures, these states are glasses. Several experimental results for supercooled water can be understood in terms of the phase diagram we present. 

\end{abstract}

\maketitle

%\section{\label{sec:level1}Introduction}
Instances of water confined to nanoscopic dimensions are ubiquitous in nature and technology.  For example, water confined to silica nanopores coated with catalyst is an efficient system for evolving oxygen, a first step towards artificial photosynthesis.\cite{Jiao:2009p7711} Another example, aquaporin pores in biological membranes confine water to channels in such a way to control water content in cells.\cite{Agre:2006p7706}  Further, silica nanopores are also used to inhibit freezing of water to enable exploration \cite{Liu:2005p2412,Zhang:2011p6972,Liu:2007p2002} of water's behavior at conditions where the bulk material would spontaneously crystalize.  The extent to which behaviors of water confined in this way -- to a long hydrophilic nanopore -- reflects behaviors of bulk water has been unknown.  Here, we address this deficiency by using general theoretical arguments coupled with molecular simulation to construct the phase diagram for water at standard and supercooled conditions as a function of temperature, $T$, pressure, $p$, and pore radius.

\begin{figure}[b]
\begin{center}
\includegraphics[width=8.2cm]{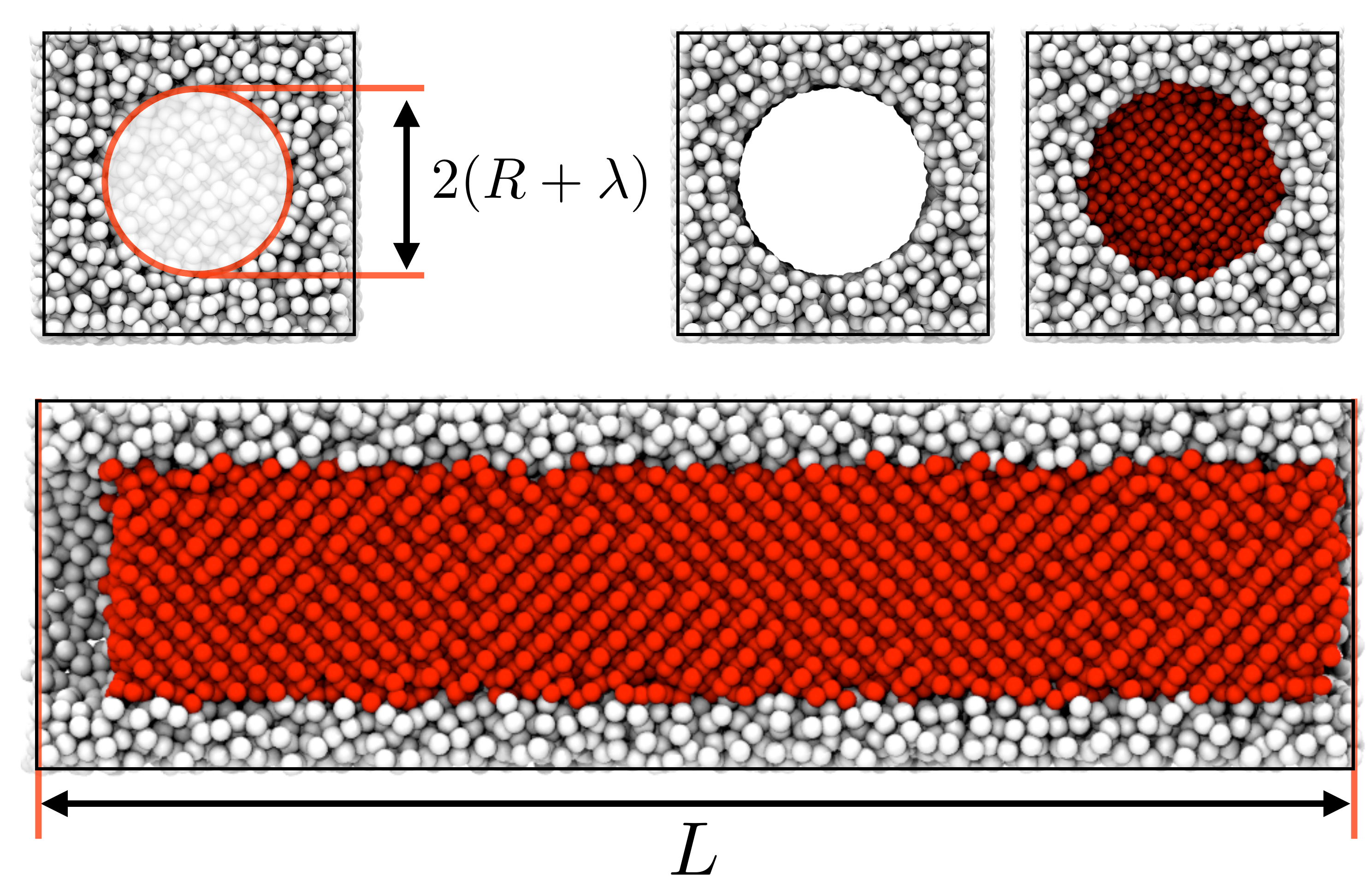}
\caption{Oxygen atoms of water (the red particles) confined in a molecular simulation to a nanopore of length $L$ with a disordered hydrophilic surface chosen to mimic silica in the material named MCM-41.\cite{Kresge:1992p7827}  The radius of the pore is $R+\lambda$, where $\lambda$ is the  thickness of an amorphous water mono-layer adjacent to the pore surface.  See text.}
\label{Fi:system}
\end{center} 
\end{figure} 

The class of systems we consider is illustrated in Fig. \ref{Fi:system}, which shows snapshots taken from a molecular simulation of water confined to a silica nanopore, details about which are given later.  This confining pore is long and narrow.  Its walls are hydrophilic, but with atoms that are in a disordered arrangement, much like a typical arrangement of oxygen atoms in liquid water, except the atoms making up the pore are frozen in place and have a slightly larger space filling size than that of water.  This static surface disorder inhibits crystal-like structures so that water adjacent to the surface is typically disordered too.  We will show that the thickness of that adjacent layer is $\lambda \approx 2.5$\,\AA, about the diameter of one water molecule.    

By considering pore radii $\Rp=R+\lambda$ that are twice 2.5\,\AA\,or larger, there can remain a significant amount of confined water that is not part of the adjacent mono-layer.  This interior water takes on ordered or disordered arrangements, depending upon temperature, pressure and pore radus.  For small enough radii, the destabilizing influence of the amorphous layer causes interior ordered water to be unstable, even at temperatures far below standard freezing temperatures. For larger radii, where ordered states are thermodynamically stable, time scales over which an ordered structure might emerge can be very long, so long that the interior water may become glass.  These are the features considered in this paper:  bulk thermodynamic stability, competing interfacial energetics and time scales to reorganize molecular structures. 

\begin{figure}[thp]
\begin{center}
\includegraphics[width=8.cm]{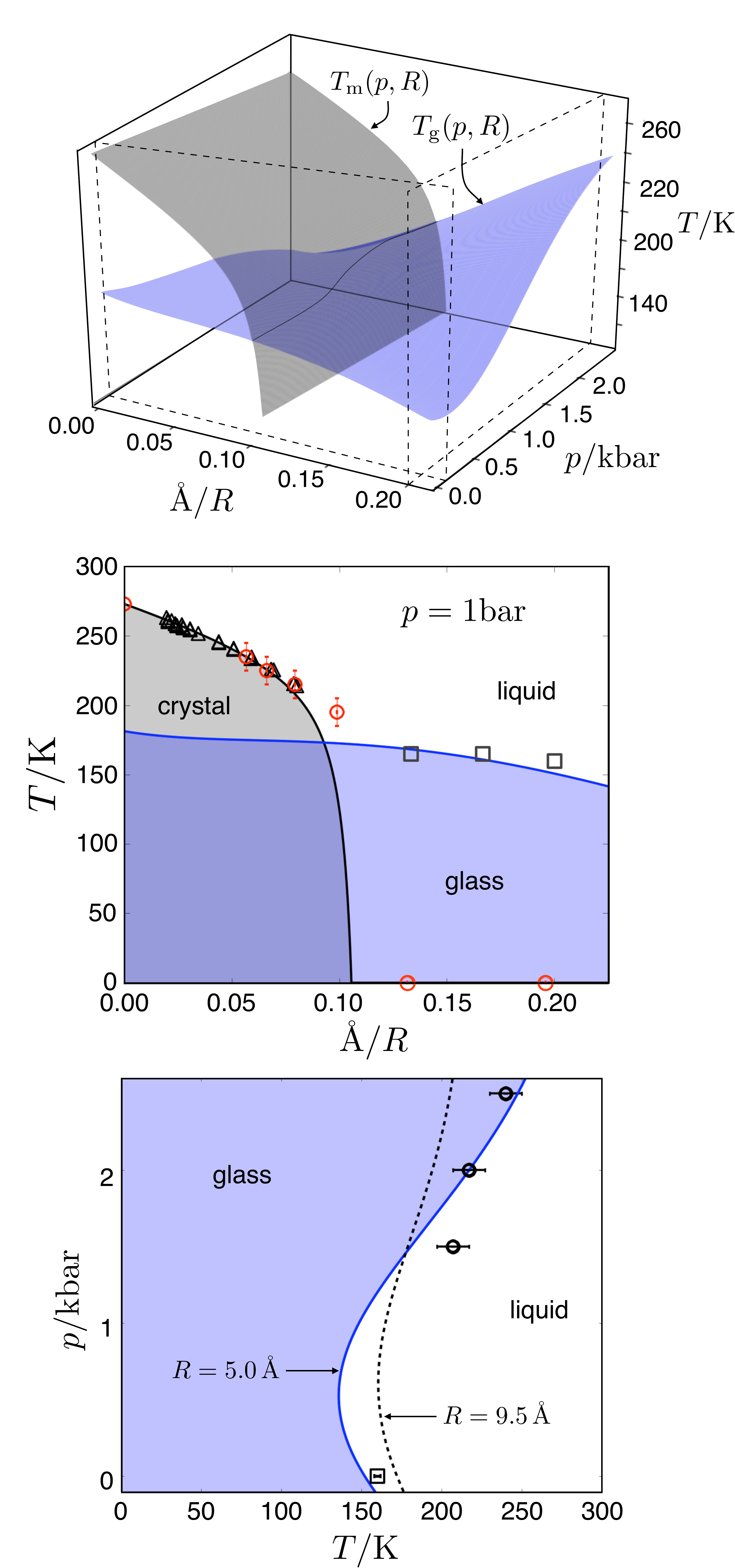}
\caption{Phase diagrams for supercooled confined water. a) Melting and glass transition temperatures, $\Tm(p,R)$ and $\Tg(p,R)$, respectively, as a functions of pressure $p$ and cylinder radius, $R$. b) Phase diagram at the constant pressure  $p= 1$\, bar. Triangular markers indicate melting temperatures measured experimentally.\cite{Findenegg:2008p1458} Circles indicate melting temperatures determined through our computer simulation, where errorbars indicate our uncertainty in $\Tm(p,R)$. Squares indicate glass transition temperatures measured experimentally with error estimates for $R$ (not shown in figure) of about $\pm$ 4 \AA .\cite{Oguni:2011p7639} c) Phase diagram for two different fixed radii $R= 5.0 \, \mathrm{\AA}$ (solid line) and $R=9.5 \, \mathrm{\AA}$ (dashed line).  Circles indicate an onset of thermal hysteresis in experimental density measurements with $R\approx 5.0 \, \mathrm{\AA}$. \cite{Zhang:2011p6972} Error bars indicate our measure of uncertainty of where hysteresis begins. The square indicates an estimate of the calorimetric glass transition for a pore of approximately the same diameter.\cite{Oguni:2011p7639} The error estimate stated in Ref. \onlinecite{Oguni:2011p7639} is smaller than the size of the symbol.}
\label{Fi:phased}
\end{center} 
\end{figure} 

\section*{Phase diagram}

Melting in a bulk macroscopic system coincides with a singularity in a free energy function.  In a bounded system, like those we consider here, the transition is smoothed or altogether removed.  Two relevant length scales associated with this behavior emerge from the microscopic theory presented in the next section.  The first, 
\begin{equation}\label{Eq:lm}
\ell_\mathrm{m} = 2\gamma / \Delta h\,\approx 0.21\, \text{nm}\, ,
\end{equation}
reflects competition between bulk energetics favoring order and interfacial energetics opposing order.  Here, $\gamma$ is the surface tension between the ordered crystal and the disordered liquid, and $\Delta h$ is the heat of fusion per unit volume.  The value, $\ell_\mathrm{m} \approx 0.21$\,nm for water, follows from the values of surface tension and heat of fusion for water-ice coexistence.\cite{NBS_steam,Granasy:2002p7516}\footnote{Reference \onlinecite{NBS_steam} gives $\Delta h \approx 3.0 \times 10^5 \,\mathrm{kJ/m^3}$, and Ref. \onlinecite{Granasy:2002p7516} gives $\gamma \approx 32 \, \mathrm{mJ/m^2}$.}  Both $\gamma$ and $\Delta h$ are pressure dependent, but the ratio $\gamma/ \Delta h$ is pressure-independent to a good approximation.\cite{TURNBULL:1950p7191}  More is said about this fact later. 

When the radius $R$ is significantly larger than $\ell_\mathrm{m}$, a melting temperature remains finite.  This temperature, $T_\mathrm{m}(p,R)$, is defined as that where the free energy of an ordered structure equals the free energy of a liquid.  According to macroscopic thermodynamics, $\Tm(p,R)$ follows a Gibbs-Thompson equation (like the Kelvin equation in the context of capillary condensation).\cite{Evans:1990p6877}  Specifically, $\Tm (p,R) \approx \Tm(p)(1-\ell_\mathrm{m}/R)$, where $T_\mathrm{m}(p)$ is the bulk melting temperature.  This approximation describing the reduction in melting temperature with increasing $1/R$ is correct to the extent that $\ell_\mathrm{m} / R\ll 1$. The melting curve for small $1/R$ shown in Fig.~\ref{Fi:phased} follows this equation. 

The second relevant length emerging from the theory manifests fluctuations that destabilize order.  Specifically, fluctuations renormalize the first length to yield
\begin{equation}\label{Eq:ls}
\ell_\mathrm{s} = \ell_\mathrm{m} / (1- T_\mathrm{s}/T_\mathrm{m})\,\approx 0.91 \, \text{nm}\,,
\end{equation}
where $T_\mathrm{s}$ stands for the temperature below which a bulk amorphous phase of water is unstable.  It is generally pressure dependent, but according to our simulation studies of one water model,\cite{Limmer:2011p6779} the ratio $T_\mathrm{s}(p)/T_\mathrm{m}(p)$ is pressure independent.  We therefore omit explicit reference to its pressure dependence in Eq.~\ref{Eq:ls}. Experimentally, it is difficult to measure $\Ts$, so in order to estimate it for water we rewrite the ratio as $\Ts/\Tm = (\Ts/T_\mathrm{\rho \, max})( T_\mathrm{\rho \, max}/\Tm)$ where $T_\mathrm{\rho \, max}$ is the temperature of maximum density at low pressure (Ref. \onlinecite{Limmer:2011p6779} uses the symbol $T_\mathrm{o}$ for that temperature). We write these ratios because we have found previously that $T_\mathrm{\rho \, max}$ represents the relevant energy scale for supercooled water thermodynamics. Therefore, we expect that for any reasonable model of water $\Ts/T_\mathrm{\rho \, max}$ will be independent of the specific choice of model. As such, it can be extracted from simulation, with which we find it to be $\Ts/T_\mathrm{\rho \, max}=0.76$.\cite{Limmer:2011p6779} The second term, $T_\mathrm{\rho \, max}/\Tm$, is a model dependent constant, often close to unity and its value for water is known experimentally to be 1.01.\cite{NBS_steam} We use that value. Therefore, we predict that $T_\mathrm{s}(1 \,\mathrm{atm})= 210$\,K for water. This prediction of a lower temperature limit to liquid stability is consistent with experimental observations of rapid spontaneous crystallization of water at 220\,K.\cite{NilssonPrivate}  In addition, it yields the value $\ell_\mathrm{s} = 0.91$\,nm cited above.

The Gibbs-Thompson correction to the bulk melting line is accurate only when order parameter fluctuations can be neglected.  These fluctuations become dominant as $R$ approaches $\ell_\mathrm{s}$.  Specifically, in the next section we derive 
\begin{equation}\label{Eq:Tm} 
 \Tm(p, R) \approx \Tm(p) \left ( 1- \frac{\ell_\mathrm{m}}{R} - \frac{\ell_\mathrm{s}^2}{8 \pi\,(R-\ell_\mathrm{s}) \, R}   \right )\, ,  
\end{equation}
for $R>R_\mathrm{c}$, where $R_\mathrm{c}$ is the positive root of the right hand side of Eq.~\ref{Eq:Tm} and is approximately equal to $ \ell_\mathrm{s}$.\footnote{$R_\mathrm{c} = \ell_\mathrm{s} \left [1/2 + \ell_\mathrm{m}/2\ell_\mathrm{s}+1/2\left  (1+1/2\pi-\ell_\mathrm{m}^2/\ell_\mathrm{s}^2\right )^{1/2}\right ]$.} For $R\le R_\mathrm{c}$, $ \Tm(p, R)=0$. This expression is graphed in Fig.~\ref{Fi:phased}.  The fluctuation contribution produces the precipitous end to the melting line near $R \approx 1$\,nm.  The comparison of data points and lines in Fig.~\ref{Fi:phased} shows that our predicted behavior of $\Tm(p,R)$ agrees well with observed calorimetry results for an order-disorder transformation of water in silica pores.\cite{Findenegg:2008p1458}  Equation~\ref{Eq:Tm} also agrees well with our molecular simulation results discussed later in this paper.    

The second surface shown in Fig.~\ref{Fi:phased}, $T_\mathrm{g}(p,R)$, is defined to be the temperature below which the structural relaxation time, $\tau$, of supercooled liquid water would be larger than 100\,s.  Nonequilibrium perturbations taking place on shorter time scales, such as cooling rates in the range of 0.1\,K/min to 1\,K/min or faster, would take the liquid out of equilibrium.  It is in that sense that $T_\mathrm{g}(p,R)$ is the glass transition temperature. To estimate its behavior, we note that $\tau(T)$ generally follows the parabolic form below an onset temperature,\cite{Elmatad:2009p7533,Elmatad:2010p7855} i.e.,
\begin{equation}\label{Eq:parab}
\log \left( \tau / \tau_\mathrm{o} \right) = J^2\, \left( 1/T \, -\,1/T_\mathrm{o} \right)^2\,,\,\,T<T_\mathrm{o}\,. 
\end{equation}
We adopt this expression together with $100\,\mathrm{s} = \tau(T_\mathrm{g}, p, R)$.  The reference time, $\tau_\mathrm{o}$, the onset temperature, $T_\mathrm{o}$, and the energy scale, $J$, are generally functions of $p$ and $R$. These functions can be determined from simulation and experiment.  

One such determination is that $\tau_\mathrm{o}$ for water is close to 1\,ps throughout the range of $p$ and $R$ we find relevant.  Accordingly, Fig.~\ref{Fi:phased} graphs
\begin{equation}\label{Eq:Tg}
T_\mathrm{g}(p,R) \approx T_\mathrm{o}(p, R) \Big / \left[1 + \sqrt{14\,}T_\mathrm{o}(p, R)/J(p,R) \right]\,.
\end{equation}
From experiment and simulation of water, we can determine functional forms for $J(p,R)$ and $\To(p,R)$.  See Methods section.  While each depends monotonically on $p$ and $R$, their systematic trends lead to the non-monotonic behavior $T_\mathrm{g}(p, R)$ illustrated in Fig. \ref{Fi:phased}.  Most notable is how the slope of $T_\mathrm{g}(p,R)$ with respect to $p$ changes from small and negative when $R$ is large to relatively large and positive when $R$ is small.  This variation in slope explains a few critical observations.

In particular, recent calorimetry experiments probing glassy relaxation in confined systems have estimated $\Tg (p,R)$ for several pore sizes  \cite{Oguni:2011p7639} at low pressures.  The results of those observations (the squares in Fig. \ref{Fi:phased}) coincide closely with our predictions for this glass transition temperature.  Furthermore, Zhang et al. \cite{Zhang:2011p6972} have, in effect, located the glass transition temperature at higher pressures through their observation of hysteretic behavior for density in nano-pores upon cooling at a rate of 0.2 K/min.  Hysteresis occurs only because the system falls out of equilibrium.  Data points from Ref.~\onlinecite{Zhang:2011p6972} (the circles in Fig. \ref{Fi:phased}) fall close to our predicted glass transition temperature line.  That reference attributes the hysteresis to something other than a glass transition, namely a hypothesized liquid-liquid transition.\cite{Poole:1992p2103}  Previous work by us casts doubt on that possibility,\cite{Limmer:2011p6779} leaving the glass transition as a plausible explanation for pore sizes as small as those reported in Ref.~\onlinecite{Zhang:2011p6972}.\cite{Soper:2011p7777,Zhang:2011p7778}  If the pore sizes were a factor of 2 larger than estimated by those authors, our phase diagram indicates that hysteresis could also reflect time scales for nucleating an ordered crystal-like material.

At temperatures below $\Tg(p,R)$, water may exist in more than one amorphous solid state.  Preparations and transitions between these amorphous states are irreversible and therefore beyond the scope of this paper.

\section*{Derivation of phase diagram}

Our approach for analyzing remnants of first-order phase transitions in bounded systems begins by choosing a general phenomenological hamiltonian for an order-parameter field parameterized with experimental data. We then perform statistical mechanical calculations for the bounded systems based upon that hamiltonian, and we test assumptions in our analysis with atomistic simulations.  Our strategy for examining the possibility of out-of-equilibrium transitions to glassy sates is to use scaling principles \cite{Elmatad:2009p7533,Keys:2011p7282} to bootstrap to the glass transition from knowledge of structural relaxation times at moderately supercooled conditions, and to use molecular simulation to test assumptions underlying that approach.

\subsection*{Equilibrium}

We consider an energy functional or hamiltonian of an order parameter distinguishing a liquid-like state from a crystal-like state.  For bulk water, the two states can be distinguished with a global order parameter like Steinhardt, Nelson and Ronchetti's $Q_6$ variable.\cite{Steinhardt:1983p1432}  Complex fields for local order parameters could be used too.  Broken symmetry for either $Q_6$ or a phase of a complex field does not occur for confined systems like those we consider here.  Therefore, we choose to distinguish liquid-like states from more ordered crystal-like states in terms of a local order field that is real.  There are many such measures suitable for our purpose.  As a specific example, our choice of order parameter could be
\begin{equation}
q(\mathbf{r}) + q_\mathrm{liq} = \sum_{i=1}^{N} q^{(i)} \, \delta(\mathbf{r} - \mathbf{r}_i)\,,
\label{eq:orderparameter}
\end{equation}
where $\mathbf{r}_i$ is the position of the $i$ oxygen among $N$ water molecules, and 
\begin{equation}
q^{(i)} = \frac{1}{4}\left( \sum_{m=-6}^{6}\,\, \Big| \sum_{\,j\in \mathrm{nn}(i)} q_{6m}^{(j)} \,\,\Big|^2   \right)^{1/2}\,,
\label{eq:q6i}
\end{equation}
with
\begin{equation}
q_{6m}^{(i)} = \frac{1}{4} \sum_{\,\,j \in \mathrm{nn}(i)} Y_{6m} (\phi_{ij},\theta_{ij})\,.
\label{eq:q6mi}
\end{equation}
Here, the sum over $j\in \mathrm{nn}(i)$ includes only the 4 nearest neighbor oxygens of the $i$th oxygen, and  $Y_{6 m} (\phi_{ij} , \theta_{ij})$ is the $\ell = 6 ,m$ spherical harmonic function associated with the angular coordinates of the vector $\mathbf{r}_i - \mathbf{r}_j$  joining molecules $i$ and $j$ measured with respect to an arbitrary external frame.  This particular order parameter is large in proportion to the concentration of water molecules with neighbors having the same orientations of neighboring bonds as does the molecule itself.  The quantity  $q_\mathrm{liq}$ is its non-zero value for the bulk liquid.  Past experience has shown that using the $\ell = 6$ spherical harmonics with 4 nearest neighbors is particularly useful for detailing local structure in water.\cite{Limmer:2011p6779}

With this or some similar order-parameter field, we choose the energy to have the following form
\begin{equation}\label{Eq:ham}
 \mathcal{H}[q(\mathbf{r})] = k_\mathrm{B} T \int_V \, \mathrm{d} \mathbf{r} \,  \left[ f( q(\mathbf{r})) + \frac{m}{2} | \nabla q(\mathbf{r}) |^2 \right]\,,
 \end{equation}
 \begin{equation}\label{Eq:f}
f(q) = \frac{a}{2}  q^2 - w q^3 + u q^4\,,
\end{equation}
where $k_\mathrm{B}$ is Boltzmann's constant, $q(\mathbf{r})$ is the deviation of the order parameter field from its uniform value for bulk liquid water, and $a = a_0(T-T_\mathrm{s})$, with $T_\mathrm{s}$ being the temperature below which the bulk liquid is unstable.  The parameters $a_0$, $w$, $u$, $m$ and $T_\mathrm{s}$ are positive constants that depend upon pressure but are independent of temperature.  They can be determined in terms of the measured properties of bulk water, all as specified below.  The zero of energy is that of the disordered amorphous material.

The most significant feature of this phenomenological energy functional is the presence of only one field, specifically one that refers to local order, and that local molecular density is explicitly absent.  We adopt this feature for two reasons.  First, our prior simulation work indicates that the reversible free energy function for condensed water has no more than one amorphous-state basin and no more than one crystal-state basin at the conditions we consider.\cite{Limmer:2011p6779}  Second, we show below that we do not need to invoke the possibility of two liquid states to explain the experimental data we set out to interpret.  In our picture, therefore, molecular density introduces no additional phase-transition-like behavior.  To the extent our picture is accurate, integrating out density therefore affects only the values of the parameters already included.   

Two other significant features of the energy functional is the truncation at fourth order in the order parameter and the neglect of inhomogeneity beyond the square gradient term.  The first feature limits our treatment to no more than two distinct reversible phases.  As noted above, we believe this limitation is acceptable for the conditions we consider.   The second feature limits our treatment to a long wavelength description of interfaces. It is the simplest description for estimating the role of surface energetics.\cite{Chaikin_book} In our simulations for the crystal-like phases, we find instantaneous domains with faceted surfaces. These features are smoothed by averaging over the disorder imposed by the pore wall. The energetics we aim to describe with the square-gradient term is for interfaces averaged over the full length of the pore. The Methods section presents tests of this idea.

Given the energy functional, the free energy, $F(T,p)$, is determined by the partition function,
\begin{equation}\label{Eq:pf}
\frac{F(T,p)}{k_\mathrm{B}T}=-\ln \int \mathcal{D}q(\mathbf{r}) \, \exp \left\{  - \mathcal{H}[q(\mathbf{r})] / k_\mathrm{B} T \right\} \,,
\end{equation}
where the dependence upon pressure, $p$, enters through the parameters in the model, as discussed above and detailed further below.

\subsection*{Mean field treatment}
Mean field theory identifies the mean value $\langle q(\mathbf{r}) \rangle$ as the function $q(\mathbf{r})$ that minimizes $\mathcal{H}[q(\mathbf{r})]$ subject to the boundary conditions imposed by the cylindrical pore.  Specifically, the free energy in this approximation is
\begin{equation}\label{Eq:mf}
F_\mathrm{MF} = \mathcal{H}[ \langle q(\mathbf{r}) \rangle]\,,
\end{equation} 
where
\begin{equation}\label{Eq:min}
\frac{\delta \mathcal{H}}{\delta \langle q(\mathbf{r}) \rangle}=0\,.
\end{equation}\\
For boundary conditions applied to solving Eq.~\ref{Eq:min}, we assume the effects of the pore are two-fold.  First, we assume the pore confines water to a cylinder of radius $R_\mathrm{p}$.  Second, we assume that disorder of the pore's confining hydrophilic surface induces liquid-like behavior in adjacent water, making 
\begin{equation}\label{Eq:bc}
\langle q(\mathbf{r}) \rangle = 0\,,\,\, \text{for} \,\, |\mathbf{r}| \ge R_\mathrm{p} - \lambda   \equiv R,
\end{equation}
where $\lambda$ is the thickness of the amorphous boundary layer.  Simulation results for various temperatures and pore radii (see Methods section) show that crystal-like domains are limited to an inner cylinder of radius $R=R_\mathrm{p} -2.5 \mathrm{\AA}$ where $R_\mathrm{p}$ is the mean distance from the center of the pore to the silica wall.  Therefore, we take $\lambda\approx 2.5 \mathrm{\AA}$ for all temperatures and pore radii. 

In this picture, surface energetics controlling the non-trivial behavior of $\langle q(\mathbf{r}) \rangle$ is determined by the interface between liquid water and ice.  The silica-water interactions are irrelevant except for producing a layer of disordered water and therefore imposing the boundary condition Eq.~\ref{Eq:bc}.  One important consequence is that the parameter $m$ is determined by the interfacial energy of ice in contact with liquid water.

For an analytical solution to Eq.~\ref{Eq:min}, we consider $R$ to be very large.  To leading order in $1/R$, the solution to Eq.~\ref{Eq:min} is that of a one dimensional interface that yields a mean field free energy per unit volume 
\begin{equation}\label{Eq:F_MF}
\frac{F_\mathrm{MF}(q)}{k_\mathrm{B} TV} = q^2 \left [ \frac{a}{2}+\frac{w}{3R}\left({\frac{2m}{u}}\right )^{1/2} - wq+ uq^2 \right ]\,,
\end{equation}
where
\begin{equation}\label{Eq:min2}
\partial F_\mathrm{MF} / \partial q=0 \,,
\end{equation} 
with $q$ being the mean-field estimate of $ \langle q(\mathbf{r})  \rangle$ for $\mathbf{r}$ at the center of the cylinder.  With numerical solutions, we have checked that terms beyond linear in $1/R$ would be significant contributors to $F_\mathrm{MF}(q)/V$ only for $R$ so small that fluctuation corrections to the mean field approximation are dominant (see below).  At that stage, corrections to Eq. \ref{Eq:F_MF} due to growing curvature are irrelevant.  

At a point where a disordered state, $q=0$, coexists with an ordered state, $q=q_\mathrm{xtl} > 0$, the mean field condition for coexistence is 
\begin{equation}\label{Eq:coex}
F_\mathrm{MF}(0) = F_\mathrm{MF}(q_\mathrm{xtl})\,,
\end{equation}
which has a solution at a temperature $T_\mathrm{MF}(p , R)$, and an entropy difference per unit volume given by
\begin{equation}\label{Eq:entropy}
 \Delta s(p,R) = - \frac{1}{V} \left[ \frac{\partial F_\mathrm{MF}(q_\mathrm{xtl})}{\partial T} \right]_{T=T_\mathrm{MF}(p,R)} \,.
\end{equation} 
In the limit $R \rightarrow \infty$, this coexistence should coincide with bulk freezing transition because it is a first-order phase transition where fluctuation effects are not important. Accordingly, we associate $T_\mathrm{MF}(p, R \rightarrow \infty)$ with the bulk freezing temperature, $T_\mathrm{m}(p)$, and $\Delta s(p, R\rightarrow \infty) \equiv \Delta s(p)$ with the entropy change between water and ice, i.e.,  $\Delta s(p) = - \Delta h/ T_\mathrm{m}(p)$. These connections to the bulk melting transition together with the corresponding mean field approximation for surface tension,\cite{Widom_book} $\gamma = \int_0^{q_\mathrm{xtl}} \left[ 2mf(q) \right]^{-1/2}\, \mathrm{d}q $, allow us to identify all relevant combinations of parameters in the model in terms of experimentally observed properties of bulk water.\footnote{The results are $a_\mathrm{o}=2\Delta h/\Tm q_\mathrm{xtl}^2$, $w=2\Delta h (\Tm-\Ts)/\Tm q_\mathrm{xtl}^3$, $u=\Delta h (\Tm-\Ts)/\Tm q_\mathrm{xtl}^4$, and $m=18\gamma^2 \Tm/\Delta h (\Tm-\Ts) q_\mathrm{xtl}^2$}  Specifically, after some algebra Eqs.~\ref{Eq:lm} and \ref{Eq:min2} yield a mean-field expression for the melting surface,
\begin{equation}\label{Eq:GT2}
 T_\mathrm{MF}(p,R)=T_\mathrm{m}(p) \left( 1- \ell_\mathrm{m}/R \right).
\end{equation}
The mean-field approximation $T_\mathrm{MF}(p,R) \approx T_\mathrm{m}(p,R)$ is identical to the macroscopic Gibbs-Thompson estimate noted earlier.

\subsection*{Role of fluctuations}
To estimate the effects of fluctuations, we evaluate $\Delta F(q) \equiv F(q) - F_\mathrm{MF}(q)$ in a Gaussian approximation. That is,
\begin{equation}\label{Eq:DF}
\Delta F(q) = -k_\mathrm{B}T\, \ln \, \int \mathcal{D}q(r) \, \exp \{ -\Delta \mathcal{H}[q(\mathbf{r})]/k_\mathrm{B}T \}\,,
\end{equation}
with
\begin{equation}\label{Eq:DH}
\Delta \mathcal{H}[q(\mathbf{r})] \,\approx\, \frac{k_\mathrm{B}T}{2}\int_\mathbf{r} \left\{ \kappa \, [\delta q(\mathbf{r})]^2 + m \big| \nabla \delta q(\mathbf{r}) \big|^2  \right\}\,,
\end{equation}
where $\kappa=a-3w q + 6u q^2$. This approximation to $\Delta \mathcal{H}[q(\mathbf{r})]$ comes from expanding  $\mathcal{H}[q(\mathbf{r})]$ through quadratic order in $\delta q(\mathbf{r}) \equiv q(\mathbf{r}) - q$.

The geometry of the system plays a role through the Laplacian in Eq. \ref{Eq:DH}. For the cylinderical boundary conditions we consider, evaluation of the Gaussian integral prescribed by Eqs. \ref{Eq:DF} and \ref{Eq:DH} can be done by using zeroth order Bessel functions with the limits of integration restricted to allow fluctuations of wavelengths up to $2\pi/R$. 

The resulting approximation to the free energy can be used to estimate the temperatures and pressures where the ordered and disordered materials have equal statistical weight, i.e., where Eq. \ref{Eq:coex} is satisfied but with $F_\mathrm{MF}(q)$ replaced with the fluctuation corrected $F(q)$.  After some algebra, we find for $R<R_\mathrm{c}$
\begin{eqnarray}\label{Eq:Final}
\frac{\Tm(p,R)}{\Tm(p)} &=& 1- \frac{\ell_\mathrm{m}}{\Rt}  \\ \notag
 & &- \frac{\ell_\mathrm{s}^2}{8\pi  (R-\ell_\mathrm{s}) R} \left [ 1 + \mathcal{O} \left (\frac{\ell_\mathrm{m}}{R} \right )^2 \right ] \, ,
\end{eqnarray}
where  we have noted the order of neglected term. This term, $\mathcal{O} \left (\ell_\mathrm{m}^2/R^2 \right )$, refer to a curvature correction that would distinguish slab and cylinder geometries. For water at the conditions we consider, the dominant contribution for small $R$ is due to $\ell_\mathrm{s}/(R-\ell_\mathrm{s})$ being large. As such, Fourier components rather than Bessel functions could have been used to diagonalize the determenent for the Gaussian integral, and equivalently, the partition function we consider is dominated by its largest eigenvalue.

The vanishing of crystal-like stability predicted in this way, where $T_\mathrm{m}(p,R) \rightarrow 0$ for $R\rightarrow R_\mathrm{c}$, is essentially a Ginzburg criterion.\cite{Chaikin_book}  The length $\ell_\mathrm{s}$ is close to but necessarily smaller than this smallest radius, $R_\mathrm{c}$,  where crystal-like states can be stable. 

\subsection*{Nonequilibrium}

\begin{figure}[b]
\begin{center}
\includegraphics[width=8.2cm]{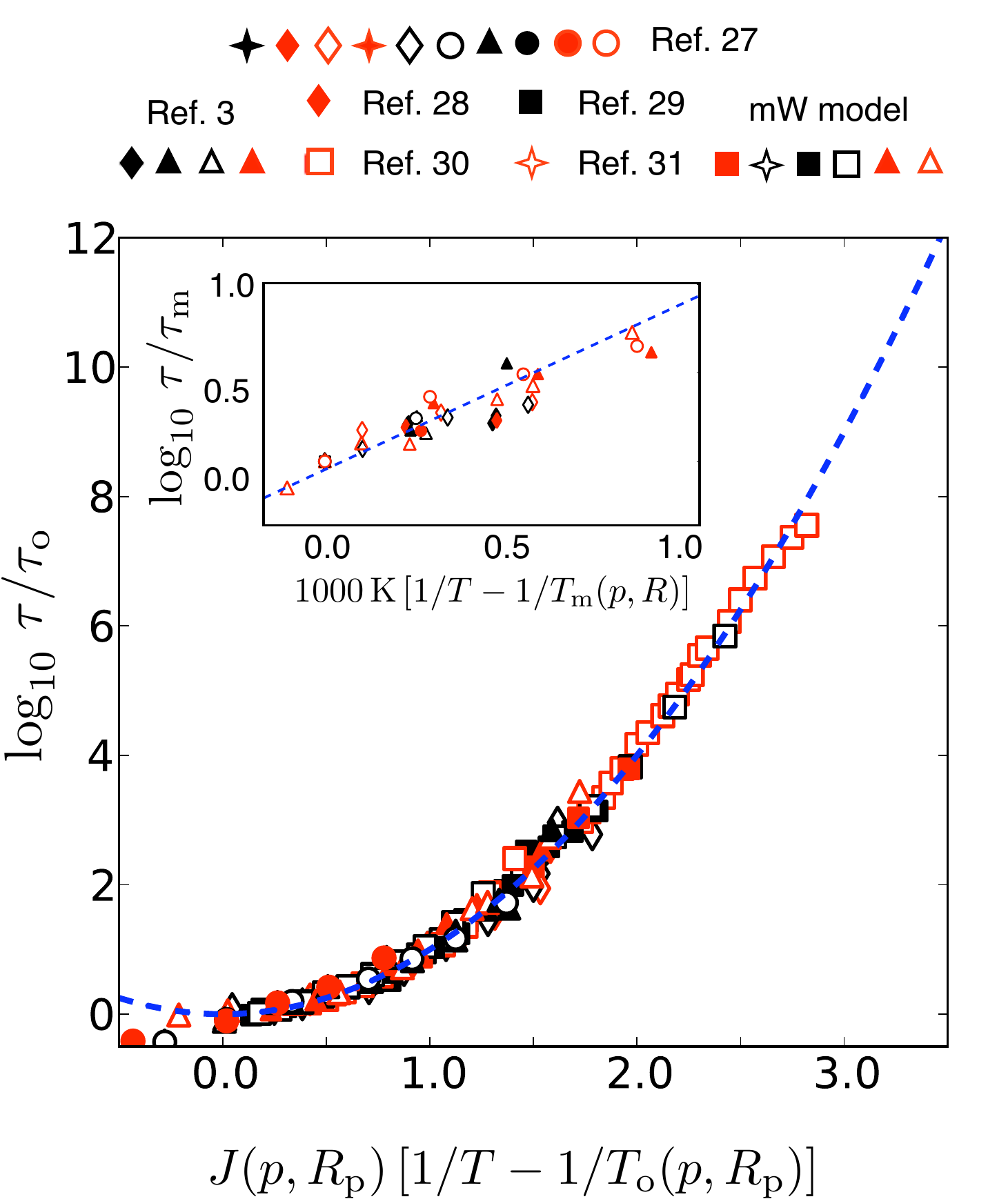}
\caption{Collapse of confined liquid water relaxation times for different pore sizes and at different external pressures. Primary graph is for $T>\Tm(p,R)$. The data is from our simulation results and experimental results.\cite{Liu:2005p2412,Liu:2006p7701,Mallamace:2006p7700,Hwang:2007p7776,Hedstrom:2007p7698,Hallett:1963p9052} Inset graph is for relaxation times of crystal-like state, i.e., $T<\Tm(p,R)$. It includes our simulation results and remaining data taken from Refs \onlinecite{Liu:2005p2412,Liu:2006p7701,Mallamace:2006p7700,Hwang:2007p7776,Hedstrom:2007p7698}.}
%illustrating the noted Arrhenious temperature dependence with an activation energy of 20 kJ/mol and attempt frequency of $\tau_\mathrm{m} = 0.5\, \mathrm{ps} \approx \tau_o$}
\label{Fi:tau}
\end{center} 
\end{figure}

To evaluate the glass transition temperature from Eq.~\ref{Eq:parab}, we must determine $\tau_\mathrm{o}$, $\To(p,\Rp)$, and $J(p,\Rp)$. These parameters control very long-time relaxation, but they can be accessed through computation and experiment that measure relatively short time behavior.~\cite{Keys:2011p7282}  The Methods section describes our handling of experimental and simulation data to obtain these parameters. For bulk water, measured relaxation times yield $\To(1\, \mathrm{atm}) \equiv \To \approx 271\, \mathrm{K}$, while the mW model used in our simulation yields $\To \approx 234 \, \mathrm{K}$; similarly, for bulk water $J(1\, \mathrm{atm}) \equiv J \approx 7.5\, \To$, while the mW model used in our simulation yields $J \approx 6.3 \,\To$.  

In creating our phase diagram, we use the real-water values for these quantities.  Nevertheless, the comparison between these quantities for real water and for the mW model give us confidence in using simulation to estimate quantities not available from experiment.  In particular, because liquid structure of mW water is virtually identical to that of real water,\cite{Molinero:2008p4576} we expect that relative dependence upon $\Rp$ can be accurately estimated with the simulation.  The dependence we find in that way for $\Rp \gtrsim 5\, \mathrm{\AA}$ is $\To(\Rp) \approx \To [ 1 + (6.0\, \mathrm{\AA}/\Rp)^2]$, and $J(\Rp) \approx J ( 1 - 4.4\, \mathrm{\AA}/\Rp)$, where $\To(\Rp)$ and $J(\Rp)$ stand for the low pressure values for $\To(p,\Rp)$ and $J(p,\Rp)$, respectively. 

For the pressure dependence of these quantities, we rely on experimental measurements of relaxation times at $\Rp \approx 7.5 \,\mathrm{\AA}$.\cite{Liu:2005p2412}  That data allows us to estimate first and second derivatives with respect to pressure, leading us to write 
\begin{equation}\label{Eq:JpR}
J(p,\Rp) \approx J(\Rp)+ 490 \, (\mathrm{K/kbar}^2)  p^2
\end{equation} 
and
\begin{equation}\label{Eq:TopR}
\To(p,\Rp) \approx \To(\Rp) - 26 \, (\mathrm{K/kbar}) \,p\,,
\end{equation} 
where $J(\Rp)$ and $\To(\Rp)$ are given in the paragraph above.

We have checked that these algebraic forms accurately extrapolate from the low-pressure values of the mW model at finite $\Rp$ to the high-pressure values for the mW model at $1/ \Rp \rightarrow 0$.  

Using these forms for the transport parameters, and a value of $\tau_\mathrm{o}=1$ ps, we can collapse experimental data and our simulation results across pressures and pore sizes. Figure \ref{Fi:tau} shows this collapse where we have restricted the data to include only equilibrium liquid relaxation, i.e., $T>\Tm(p,R)$. Figure.~\ref{Fi:tau} includes data from both experiment \cite{Liu:2005p2412,Liu:2006p7701,Mallamace:2006p7700,Hwang:2007p7776,Hedstrom:2007p7698,Hallett:1963p9052} and from our simulation study. While external pressures can be accurately controlled and reported, errors in pore sizes are large.\cite{Mancinelli:2009p7409} The Methods section discusses estimates of $\Rp$ from experimental data. 

Previous simulation\cite{Gallo:2010p7400} and experiment \cite{Liu:2005p2412,Liu:2006p7701,Mallamace:2006p7700,Hwang:2007p7776} studies have indicated that  confined liquid water undergoes an abrupt crossover in the temperature scaling of its relaxation time. This crossover is a manifestation of a transition between the liquid and crystal-like regimes, which we turn to now. 

\begin{figure}[hb!]
\begin{center}
\includegraphics[width=6.2cm]{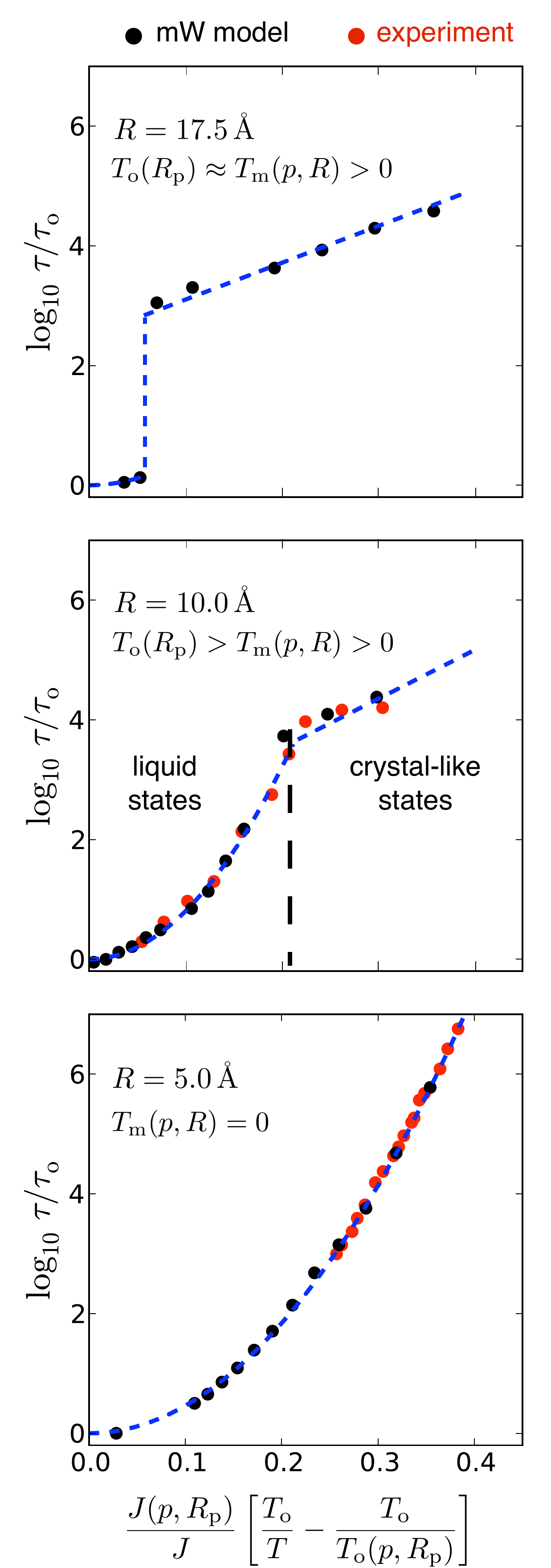}
\caption{Transport behavior for different pore sizes as indicated. Data are from simulation results (black points) and experimental results (red points). Lines are predictions based on our phase diagram and scaling relations.  Vertical dashed line in the middle panel locates the boundary between liquid and crystal-like states, i.e., where $T=\Tm(p,R)$.}
\label{Fi:crossover}
\end{center} 
\end{figure}

\section*{Relaxation of Crystal-like States}

Molecular motion of crystal-like states in confinement takes place preferentially near the water pore interface, where the molecules are locally disordered. Like defect motion in a bulk crystal, though with a smaller barrier due to the presence of the interface, the temperature dependence of such motion is expected to be Arrhenious. The inset of Fig.~\ref{Fi:tau} shows experimental and simulation data for the average time for a particle to displace one molecular diameter as a function of temperature at conditions where water in the core is ordered. Plotted is experimental data for $T<\Tm(\Rt)$ and simulation results for $R=$ 17.5 \AA.  We find that this motion is activated with a barrier of approximately 20 kJ/mol, and an attempt frequency 1/$\tau_\mathrm{m}\approx$\,2 ps$^{-1}$. There is negligible dependence on radius of confinement within the range considered. 

By combining the information of the phase diagram with our understanding of the mobility in each state we can predict the observed equilibrium behavior of the relaxation time. We find that there are three different pore size regimes, each with a distinct temperature dependence of $\tau$. These regimes are highlighted in Fig.~\ref{Fi:crossover}. First, for larger pores, $R>2\ell_\mathrm{s}$, the onset to glassy dynamics is close to $\Tm(p,R)$, therefore an equilibrium measurement should show little temperature dependence for $T>\Tm(p,R)$ and an Arrhenious temperature dependence for $T<\Tm(p,R)$ reflecting the relaxation behavior of the crystal-like states. For smaller pores close to but larger than $\ell_\mathrm{s}$, the onset temperature is greater than $\Tm(p,R)$, therefore an equilibrium measurement should show parabolic temperature dependence for $T>\Tm(p,R)$, and a crossover to Arrhenious behavior for $T<\Tm(p,R)$. For very small pores, $R<\ell_\mathrm{s}$ (but still larger than a molecular diameter), $\Tm(p,R)=0$ therefore an equilibrium measurement should show parabolic temperature dependence for $T<\To$. Figure \ref{Fi:crossover} shows that each of these regimes are observed both in simulation and in experiment. 

We are not the first to suggest that the abrupt crossover in relaxation might be linked to crystallization.\cite{Soper:2012p7879} Some may have disregarded this possibility due to the absence of a freezing peak in the heat capacity, measured by differential scanning calorimetry. Our analysis shows that the absence of this peak is due to the pore size being close to $\ell_\mathrm{s}$. When $\Rt \approx \ell_\mathrm{s}$, the ordering transition is smeared due to large structural fluctuations. As a consequence, there will be no sudden heat release. This explanation is consistent with a recent differential scanning calorimetry study that observed only partial crystallization for a pore size $\Rp= 10.5$, with the accompanying heat capacity peak being of the order of the magnitude of the maximum liquid state heat capacity.\cite{Oguni:2011p7639} 

Experimentally determined vibrational density of states for confined supercooled water differs significantly from that of bulk ice, even at points in the phase diagram where we predict the presence of crystal-like behavior.  This difference in density of states is expected because the domain of crystal-like behavior in the confined system is surrounded by a pre-melting layer, which in turn is surrounded by a layer of complete disorder.  These layers, discussed in the Methods section, encompass a significant fraction of the total system, a fraction that grows with decreasing pore size.  Further, even away from the disordered pore wall, crystal-like behavior in confinement exhibits a high concentration of stacking faults,\cite{Moore:2012p9060} which will further modify the density of states.

\section*{Methods}

\subsection*{Molecular simulation model}

The molecular dynamics simulations used to test our theoretical approximations and estimate the magnitudes of some differential changes employ the mW model of water.\cite{Molinero:2008p4576} Recently proposed by Molinero and Moore, this model has proven to yield a good description of water in the liquid phase,\cite{Moore:2009p248,deLaLlave:2010p4935,Xu:2010p6876}, it reproduces many structural transformations seen in experiment and in other models of water (including freezing into an ice-like structure),\cite{Kastelowitz:2010p6846,Li:2011p8964,Moore:2010p1923,Limmer:2011p6779} and it exhibits the characteristic thermodynamic and dynamic anomalies of water (i.e., density maximum, heat capacity increase, diffusion maximum, and so forth).\cite{Molinero:2008p4576,Limmer:2011p6779}

To model the hydrophilic pores of MCM-41-S we have followed a procedure similar to that found in Ref.~\onlinecite{Moore:2010p1872}. The pore configurations are obtained by quenching a high temperature liquid configuration of silica. To create the cylindrical geometry, we extract from the simulation box all atoms whose centers lie within a circle of radius $\Rp = [(x_i-x_c)^2+(y_i-y_c)^2]^{1/2}$ where $(x_c,y_c)$ is the center of the simulation box and $(x_i,y_i)$ is the coordinate vector for particle $i$. The remaining atoms are tethered to their initial conditions by a spherically symmetric harmonic potential with a spring constant, 50 kcal/mol \AA$^2$. This procedure yields a mean surface roughness for the pore walls in good agreement with that estimated from experiment on MCM-41-S materials.\cite{Moore:2010p1872} We have considered pore sizes in the range $R = 5.0\,$ \AA - $17.5\,$ \AA. All pores are length $L=220$ \AA \, to approach a regime where $ L >\Rt$.  For initializing the combined water, pore system, water molecules are arranged in a hexagonal ice lattice at a density of 0.98 $\mathrm{g/cm}^3$ and placed within the pore with the crystallographic c-axis parallel to the length of the pore. 

 All of the molecular dynamics trajectories were propagated using the LAMMPS package \cite{Plimpton:1995p3851} and a Nose-Hoover thermostat with constant number of particles, $N$, volume, $V$, and temperature $T$.  We fill approximately 90\% of the length of the pore.  Water organizes spontaneously with an interface separating the remaining 10\% empty pore from the condensed phase (either liquid or crystal-like).  With this procedure, we simulate the condensed material at a low pressure (effectively $p \approx 0$) in coexistence with vapor.  

We have adopted an interaction potential for a single site model of silica that is the same form as the mW model, but we have rescaled the interaction strength and particle diameter. Compared to the parameters used in the mW model, these are $\epsilon_\mathrm{silica}/\epsilon_\mathrm{mW} = 1.15 $ and $\sigma_\mathrm{silica}/\sigma_\mathrm{mW} = 1.05$. The increased interaction strength ensures a hydrophilic surface and the increased particle size frustrates local favored structures.

\begin{figure}[t]
\begin{center}
\includegraphics[width=7.6cm]{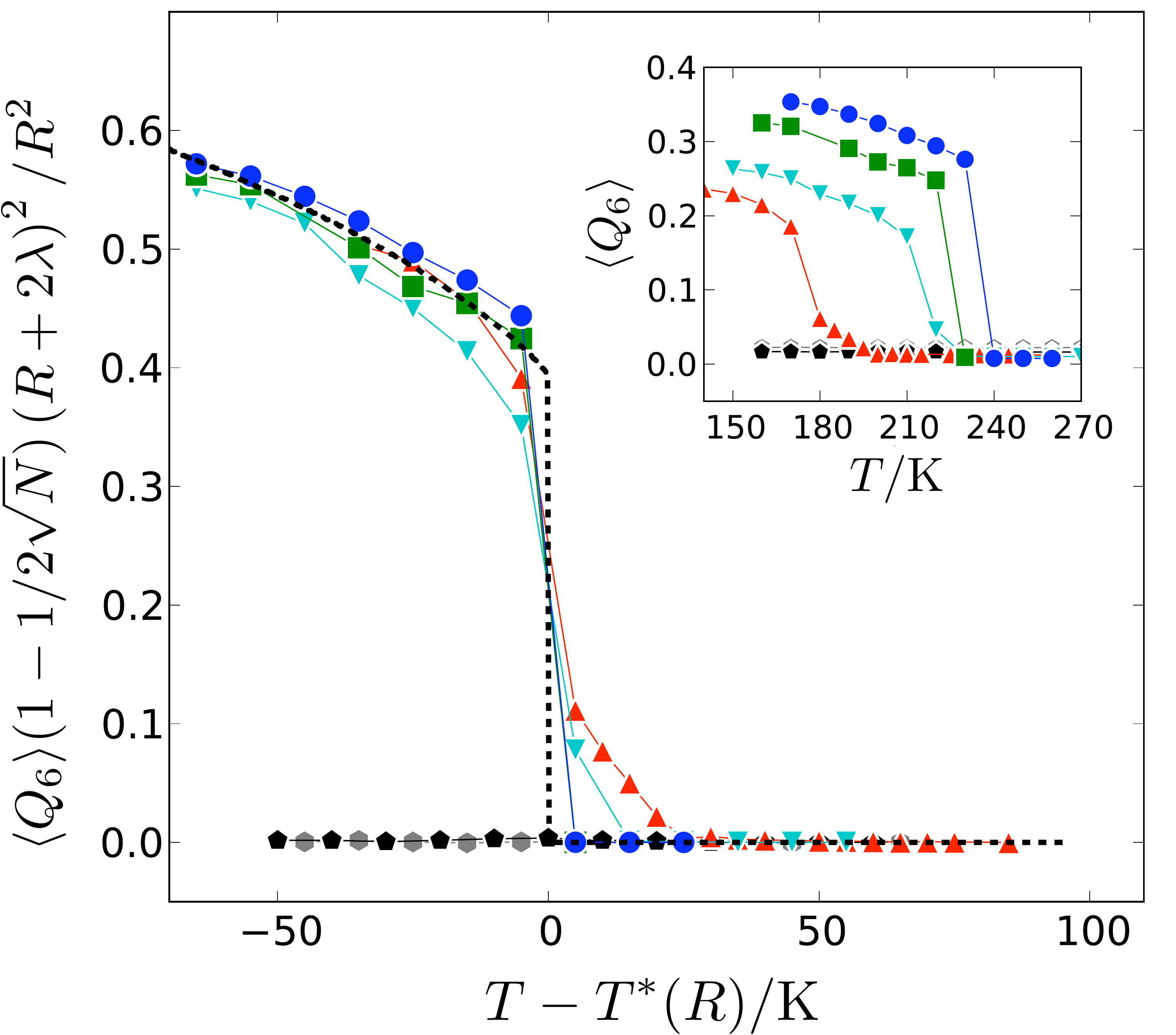}
\caption{Average value of the global orientational order parameter, $\langle Q_6 \rangle$, as a function of temperature for different pore sizes. In the main figure, $\langle Q_6 \rangle$ is rescaled and plotted as a function of $T-T^{*}(R)$, where $T^{*}$ is the temperature at which $\mid d \langle Q_6 \rangle / dT\mid$ is maximal. Inset shows the same, but not rescaled. Different markers correspond to different pore size systems, with $\Rp=\, 2.5 \, \mathrm{\AA }+ R =20.0 \, \mathrm{\AA } \, (\mathrm{blue} \, \bigcirc),17.5 \, \mathrm{\AA } \, (\mathrm{green } \, \square), 15.0\, \mathrm{\AA }  (\mathrm{cyan} \, \triangle)$, $12.5 \, \mathrm{\AA } \,(\mathrm{red} \, \triangledown)$, $10.0 \, \mathrm{\AA }  \,(\mathrm{grey} \, \pentagon)$ and $7.5$ \, \AA   $\,(\mathrm{black} \, \hexagon)$.} \label{Fi:qvt}
\end{center} 
\end{figure}

\begin{figure}[bh]
\begin{center}
\includegraphics[width=7.2cm]{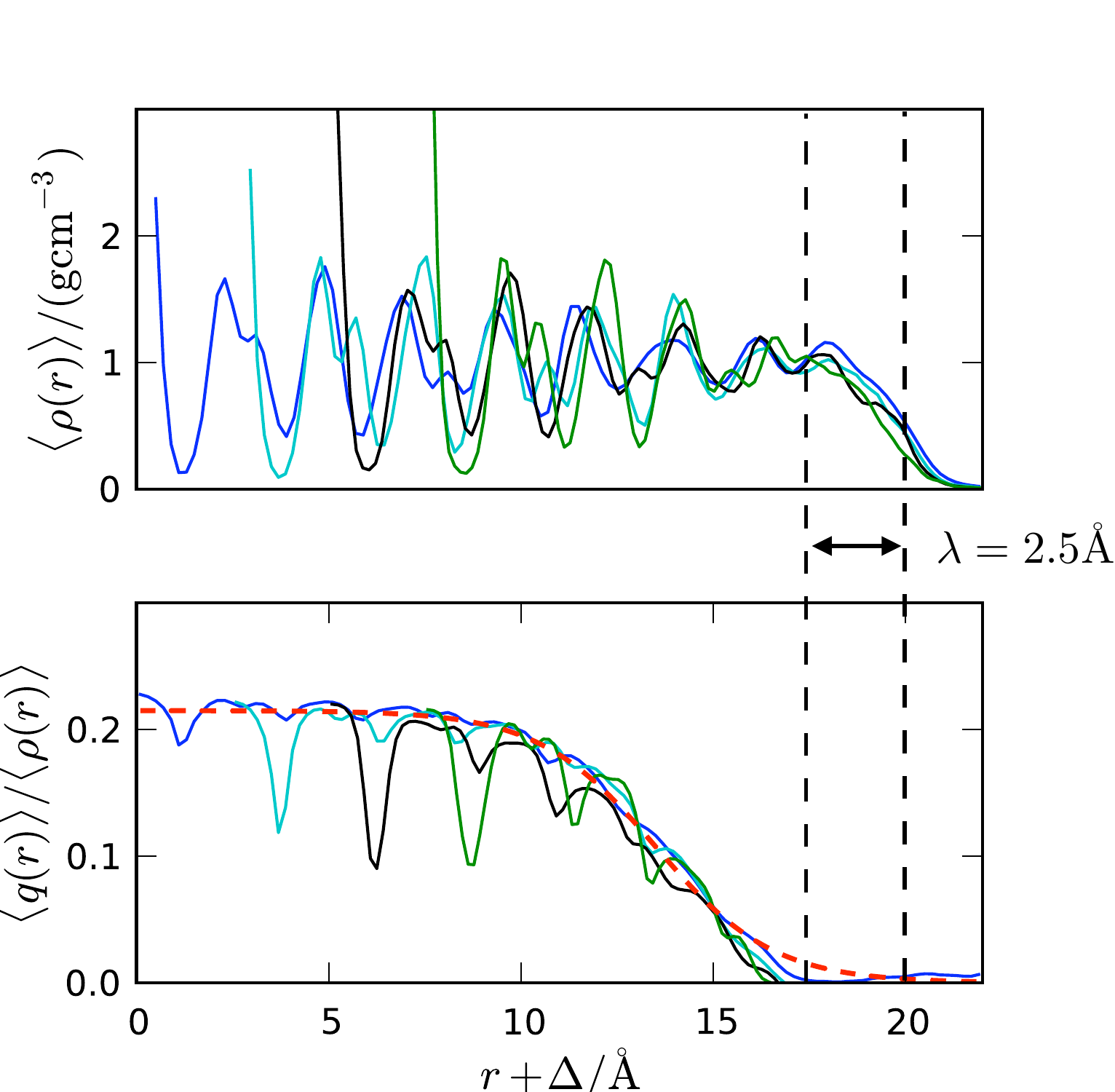}
\caption{Mean number density, $\langle \rho(r) \rangle$, and orientational order density, $\langle q(r) \rangle$, for water confined to cylindrical pores. $r$ is the radial position of the pore system. Different color solid lines are our mW model simulation results for different pore size systems, with $\Rp=20.0\, \mathrm{(blue)},17.5\, \mathrm{(cyan)},15.0\, \mathrm{(black)}$ and $12.5 \, \mathrm{(green)}$\ \AA. All are computed at $T\approx \Tm(p,R)$ and $p\approx0$. The red dashed line is the prediction from the mean field theory, Eq.13, using parameters for $\mathcal{H}[q(\mathbf{r})]$ found with the mW model. $\Delta = 20 - R_\mathrm{p}$ is used to shift $r$ to facilitate comparison of different radii pores. Notice that the order remains absent for density within $\lambda=2.5\,\mathrm{\AA}$ of the pore wall. }
\label{Fi:system2}
\end{center} 
\end{figure}
\subsection*{Determination of $\Tm(R)$ from simulation}

In order to determine the low pressure melting temperature in confinement we calculate the temperature dependence of the orientational order parameter $Q_6$ defined as 
\begin{equation}
Q_6 =\frac{1}{N} \left ( \sum_{m=-6}^6 \sum_{i,j}^N q_{6m}^{i} q_{6m}^{j*} \right )^{1/2}\, ,
\end{equation} 
following Ref.~\onlinecite{Steinhardt:1983p1432}. Rather than a local measure of order, we use this global measure of crystallinity to be sure we are distinguishing crystal from liquid.   Figure~\ref{Fi:qvt} shows the mean value of orientational order parameter, $\langle Q_6 \rangle$, as a function of temperature for six pore radii, $\Rp$, ranging between 20.0 \AA \, and 7.5 \AA.  For the range of temperatures we consider, pores with $\Rp \le 12.5 \,$ \AA \, never show pseudo long range order. Pores with $\Rp \ge 12.5 \,$ \AA \, do show pseudo long range order. The presence of the amorphous interface ensures that $\langle Q_6 \rangle$ converges relatively quickly in comparison to the bulk where large nucleation barriers would separate the ordered and disorder states at coexistence. 

Apart from shifting the coexistence temperature, the pore radius changes the maximum value $Q_6(T)$ obtains in the ordered state. This is due to the increased weight that the amorphous boundary layer has on the volume integral for decreasing $R$ and fixed thickness $\lambda$. The curves calculated for different $R$ can be collapsed by multiplying $Q_6$ by a scaling factor. This factor accounts for the total volume available within pore compared to that for crystal-like states, $(R+2\lambda)^2/R^2$, and subtractes the value for a disordered system of $N$ water molecules.

\begin{figure*}[t]
\begin{center}
\includegraphics[width=17cm]{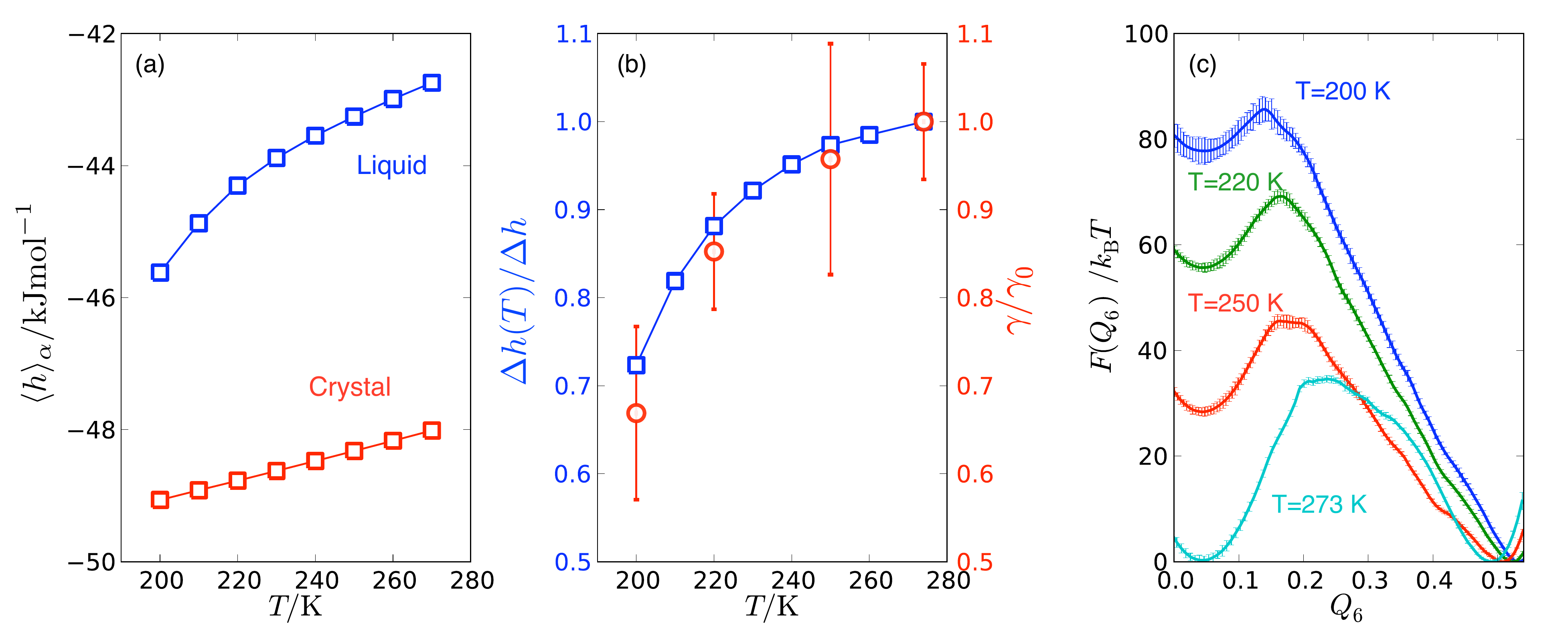}
\caption{Validation of the Turnbull relation, $\gamma / \Delta h \approx \mathrm{const}$ for the mW model. (a) Temperature dependence of the mean enthalpy in the bulk liquid and crystal states. (b) Liquid-crystal surface tension, $\gamma$, and the enthalpy of fusion, $\Delta h$, divided by their values at coexistence, $\Tm=274$ K. (c) Free energy computed for a $N=216$, $p=1$ atm bulk system illustrating data with which coexistence and surface tension is determined.}
\label{Fi:tension}
\end{center} 
\end{figure*}

\subsection*{Disorder width, curvature and pre-melting layer}
Figure 6 shows the profiles for mean density, $\langle \rho(r) \rangle$, and mean order parameter, $\langle q(r) \rangle$, as defined in Eq.~\ref{eq:orderparameter}. These curves are obtained from the mW model by simulation, and from our theory by solving Eq.~\ref{Eq:min} with the parameters appropriate for the mW model.  The thermodynamic conditions considered are where the crystal-like state is stable.  Several pore diameters were studied, and the illustrated results are typical.  In each case, the simulation yields a disordered layer of non-zero particle density and of thickness $\lambda = 2.5 \, \mathrm{\AA}$ adjacent to the pore wall.  This thickness of the disorder layer is in good agreement with the value inferred from fitting Eq.~\ref{Eq:GT2} to experimental data.\cite{Findenegg:2008p1458}

The simulations also show oscillations in both the density and the order parameter.  These oscillations reflect the size of the particles in the simulated model.  By construction, the square-gradient theory does not contain these oscillations.  Nevertheless, the rise in the mean order parameter from its disordered value at the wall to its crystal-like value in the center of the pore is consistent with those of the simulation when coarse grained over a particle diameter.  The general agreement of the profiles calculated with our molecular dynamics simulations with those calculated neglecting curvature corrections indicates those corrections are small. 

The amplitude of the oscillations in the mean order parameter obtained from the simulation results are relatively small, typically 10\% of the mean, except at the very center of the pore where statistics is unreliable.  Away from the center, the oscillations are especially small in comparison to those that would be found in an ordered crystal.  The amplitudes are diminished from those of a crystal due to the average over disorder along the length of the tube.

The width of the interface exhibits a slight temperature dependence. For larger pores, however, the situation changes.  As the radius grows beyond the conditions treated here, the coexistence temperatures will tend towards the bulk melting temperature.  A pre-melting layer between the disordered surface and the crystal will then become large and strongly sensitive to temperature.\cite{Li:2007p9167}  For macroscopic systems, this pre-melting width diverges as $T$ approaches the melting temperature.   With the equations we use in our theory, this behavior is isomorphic to a liquid-vapor wetting transition.  It is is a general behavior accompanying any first-order transition with appropriate boundary conditions.\cite{Widom_book}

The theory we have employed to describe pre-melting in a finite system would not only seem easily generalizable to treating pre-melting profiles of bulk ice in contact with its vapor or liquid,\cite{Sazakia:2012p9222} it would also seem applicable to stability and thermodynamics of nano-clusters of ice,\cite{Johnston:2012p9090} and to nucleation of ice on atmospheric aerosols.\cite{Kolb:2010p9135} It might also be generalizable to describe ordering of water in cold micro emulsions like those recently considered by Tanaka and co-workers.\cite{Murata:2012p9264}   Indications of order-disorder phenomena occurring in the finite water-rich domains of those systems have been interpreted in terms of a doubtful liquid-liquid transition in supercooled water.  Based upon what we have derived in this paper, we believe a more natural explanation of Tanaka's observations will be found in terms ice-water equilibrium and  the effects of confinement on that phase equilibrium

\subsection*{Turnbull relation}
To test the applicability of Turnbull's $\gamma/\Delta h \approx \mathrm{constant}$, we have calculated the surface tension and enthalpy of fusion as functions of temperature.\cite{TURNBULL:1950p7191} See Fig.~\ref{Fi:tension}. Here, $\Delta h(T_\mathrm{m}) = 5.4$ kJ/mol and $\gamma (T_\mathrm{m}) = 35.3$ mJ/m$^2$.

To make that figure, we have determined the enthalpy of fusion by calculating the average enthalpy density differences at coexistence, $\langle h \rangle_\mathrm{liq} -\langle h \rangle_\mathrm{xtl}$. Similarly, we have determined the surface tension by calculating the free energy as a function of $Q_6$, using the umbrella sampling procedure described in Ref.~\onlinecite{Limmer:2011p6779} for $N=216$ particles at a constant pressure, $p=1$\, atm. See Fig. \ref{Fi:tension}. The surface tension is then obtained by taking the difference between the free energy at the top of the barrier and at its stable coexisting basins.  Specifically, $\gamma = \Delta F(Q_6) / L^2$ where $\Delta F$ is the interfacial free energy calculated by first preforming a Maxwell construction to place the system at coexistence at the different temperatures, and $L=(N/\rho)^{2/3}$.\cite{BINDER:1982p7265} This procedure is exact in the limit of $N\rightarrow \infty$.  We have checked that we closely approach the limiting value by studying several system sizes up to $N=1000$ particles.  This surface tension is an effective surface tension obtained by integrating over all distinct crystallographic faces and agrees well with that obtained from a recent nucleation study.\cite{Li:2011p8964}

\subsection*{$\Tm(p,R)$ for water and the mW model}
For our calculations here we assume that the bulk melting line can be accuratly approximated by $\Tm(p) = \Tm[1 - p C+\mathcal{O}(p^2)]$. The coefficient $C$ is related to heat of fusion and the change in volume between water and ice determined at ambient pressure, as derived through the Clapeyron equation. For water $C=0.026\,\mathrm{kbar}^{-1}$  (Ref. \onlinecite{NBS_steam}) and for the mW model $C=0.01\,\mathrm{kbar}^{-1}$.\cite{Molinero:2008p4576} This difference in slope between the mW model and real water is due to the mW model over estimating the density of Ice Ih.\cite{Molinero:2008p4576} However, by defining a pressure scale in units of $C$, the equation of state of water and the mW model can be related. 

The mW model we calculate $\ell_\mathrm{m}$ and $\ell_\mathrm{s}$  to be  2.40 \AA \, and 8.16 \AA \, respectively. These are slightly different than what we find for real water, and the differences account for the differences between our predicted melting line for real water and the calculated melting temperature of the mW model for $1/R=0.1\,\mathrm{\AA}^{-1}$. See Fig.~\ref{Fi:phased}. The melting lengths for both the mW model and experiment agree with previously reported values based on fitting melting data to Eq.~\ref{Eq:GT2}.\cite{Findenegg:2008p1458,Moore:2010p1872}

\begin{figure}[t]
\begin{center}
\includegraphics[width=8.5cm]{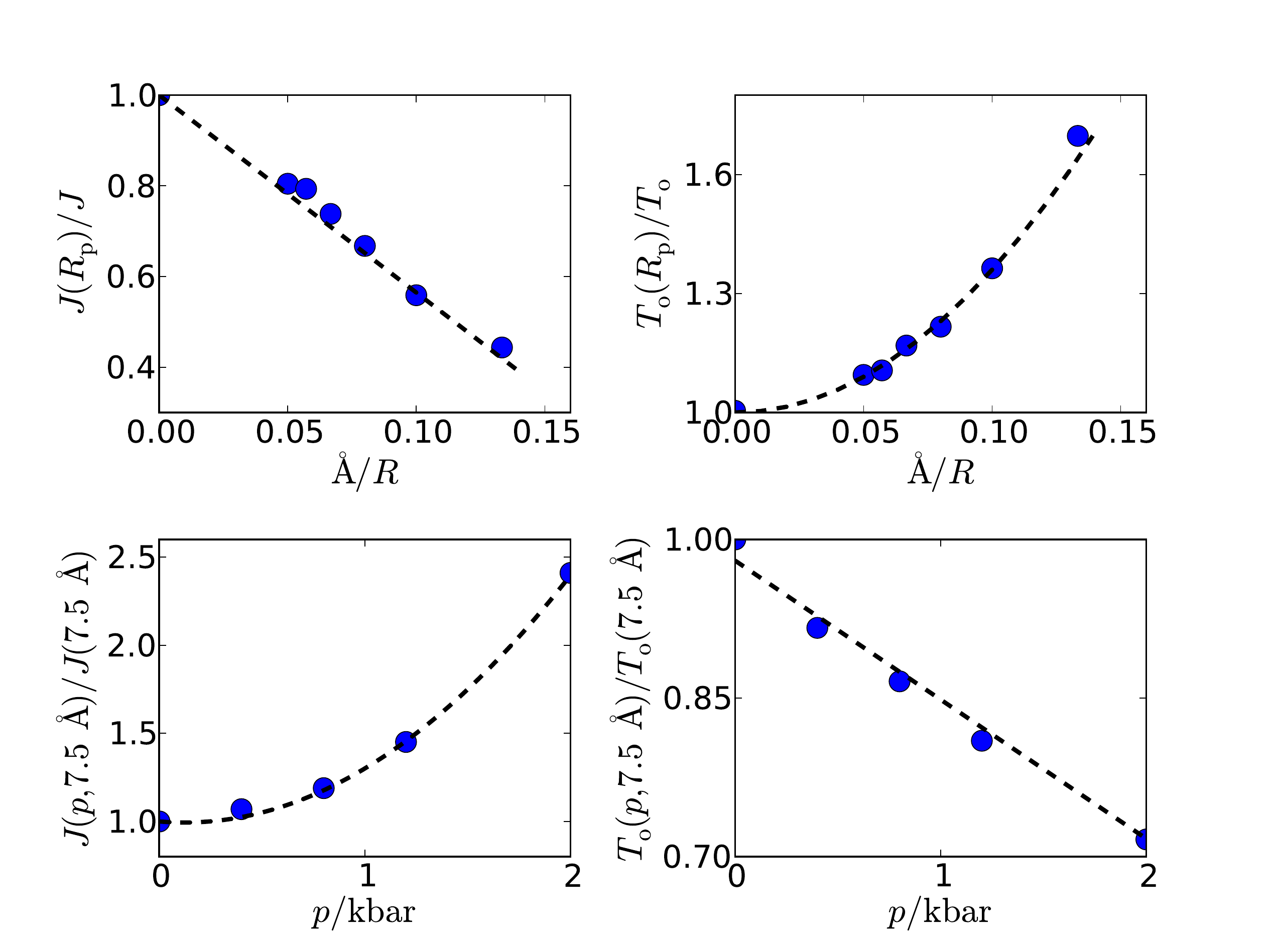}
\caption{Onset temperature, $\To(p,R)$ and energy scale, $J(p,R)$. The top two graphs are low pressure data found from simulations of the mW model. The bottom two graphs are high pressure data found from fitting experimental results of Ref.~\onlinecite{Liu:2005p2412}. In that case, the experiments report $\Rp=7.5\mathrm{\AA}$.The dashed lines are the curves obtained with Eqs.~\ref{Eq:JpR} and \ref{Eq:TopR}.}\label{Fi:scale}
\end{center} 
\end{figure}

\subsection*{Transport regression analysis}

We preformed a regression analysis on the algebraic forms used for $J(p, \Rp)$ and $\To(p,\Rp)$, Eqs.~\ref{Eq:JpR} and \ref{Eq:TopR}. See Fig. \ref{Fi:scale}. The top two panels concern the dependence upon $\Rp$, and the data for relaxation times is obtained from our molecular dynamics simulations of the confined mW model.  The bottom two panels concern the pressure dependence, and the data for relaxation times is taken from experiments on confined water with $\Rp=7.5$ \AA.\cite{Liu:2005p2412} The dashed lines are our algebraic fits, where the correlation coefficients indicate a certainty of 1\% or better. Relaxation times for the molecular dynamics simulations were determined by calculating the mean time for a particle to displace one diameter,

\subsection*{Determination of $\Rp$ from experimental data}

While nominal nanopore radii are routinely reported in the literature, it is difficult to obtain a reliable estimate of $\Rp$ for $\Rp<$ 1 nm. Different techniques yield a range of different sizes.\cite{Mancinelli:2009p7409} Most commonly, pore sizes are inferred from a Barrett-Joyner-Halenda (BJH) analysis.\cite{Barrett:1951p7468} This method amounts to measuring a nitrogen absorption isotherm, and is thus an indirect measure of size. Mancinelli et al have demonstrated that this method can yield significant errors, up to 200 \%. For example, using refined neutron scattering data and mass balance calculations, Mancinelli et al. have estimated a likely range of pore sizes for the system studied by Ref. \onlinecite{Liu:2005p2412} to be between $7.5\mathrm{\AA}$ - $12.6\mathrm{\AA}$,\cite{Mancinelli:2009p7409} while the BJH method yields $7.5\mathrm{\AA}\pm 2\mathrm{\AA}$.  

Using the bounds provided by Ref.~\onlinecite{Mancinelli:2009p7409} as reliable estimates of possible errors, we find that we can collapse the experimental transport data, but such a collapse cannot be obtained within the errors reported from the BJH method. The pore sizes inferred from this collapse indicates that the BJH method systematically underestimates pore sizes.\cite{Zhang:2011p7778} Previous studies claiming to study the same pore sizes have observed widely different behavior. For instance, Ref. \onlinecite{Liu:2005p2412} report a pore radii $\Rp=7.5\mathrm{\AA}$, and measure a relaxation time that is never larger than 10's of nanoseconds. Reference \onlinecite{Oguni:2011p7639} in one experiment also report using a pore of radii $\Rp=7.5\mathrm{\AA}$ and measure thermal signatures indicative of a glass transition implying relaxation time on the order of seconds. In light of our results detailing the different transport regimes that can occur for slightly different pore sizes, the implications of the errors associated with the reported values of the pore size become significant. 

\begin{acknowledgments} We are grateful to Christopher Bertrand, Pablo Debenedetti, Aaron Keys, Valeria Molinero, Suriyanarayanan Vaikuntanathan and Yang Zhang for their reading and suggestions on an earlier draft of this paper.  Work on this project in its early stages was supported by the Director, Office of Science, Office of Basic Energy Sciences, Materials Sciences and Engineering Division and Chemical Sciences, Geosciences, and Biosciences Division of the U.S. Department of Energy under Contract No. DE-AC02-05CH11231. In its final stages, it was supported by the Helios Solar Energy Research Center under the same DOE contract number.
\end{acknowledgments} 

%\bibliography{ref_paper,ref}
%\end{thebibliography}

% Produces the bibliography via BibTeX.
\end{document}